# Penalized-likelihood PET Image Reconstruction Using 3D Structural Convolutional Sparse Coding

Nuobei Xie[†], Kuang Gong[†], Ning Guo, Zhixin Qin, Zhifang Wu, Huafeng Liu*, Quanzheng Li*

*Abstract*— **Positron emission tomography (PET) is widely used for clinical diagnosis. As PET suffers from low resolution and high noise, numerous efforts try to incorporate anatomical priors into PET image reconstruction, especially with the development of hybrid PET/CT and PET/MRI systems. In this work, we proposed a novel 3D structural convolutional sparse coding (CSC) concept for penalized-likelihood PET image reconstruction, named 3D PET-CSC. The proposed 3D PET-CSC takes advantage of the convolutional operation and manages to incorporate anatomical priors without the need of registration or supervised training. As 3D PET-CSC codes the whole 3D PET image, instead of patches, it alleviates the staircase artifacts commonly presented in traditional patch-based sparse coding methods. Moreover, we developed the residual-image and order-subset mechanisms to further reduce the computational cost and accelerate the convergence for the proposed 3D PET-CSC method. Experiments based on computer simulations and clinical datasets demonstrate the superiority of 3D PET-CSC compared with other reference methods.**

*Index Terms*—**Positron emission tomography, convolutional sparse coding, 3D image reconstruction, anatomical prior, multi-modality.**

## I. INTRODUCTION

AS an irreplaceable tool of functional imaging, positron emission tomography (PET) is widely adopted in clinical diagnosis for oncology [1], neurology [2] and cardiology [3]. By collecting photons emitted from a specific tracer, PET manages to recover the physiology-based tracer distribution map in *vivo* [4]. Granted PET is proved sensitive in the molecular level, it is still inferior in recovering high-resolution details when compared with computed tomography (CT) and magnetic resonance imaging (MRI). Numerous efforts have been devoted to improving PET resolution and reducing the noise level through denoising/reconstruction approaches. For the past decades, most of the research focus on incorporating specific penalties or image priors into the iterative reconstruction framework to improve PET image quality, e.g.

total variation (TV) [5], patch-based edge-preserving regularization [6], nonlocal mean-based weight [7], and the kernel method [4].

Accompanied with the rapid adoption of hybrid PET/CT and PET/MRI systems as well as the development of machine learning methodologies, PET image quality can be improved by the utilization of CT/MRI images, through convolutional neural network (CNN) [8]–[10] and dictionary learning [11]–[18] approaches. For CNN methods, the applications are limited by two concerns: firstly, these methods heavily rely on the matching between PET and CT/MR images, making the time-consuming registration indispensable; secondly, the CNN methods require enormous training data, which is not easy to obtain/process, especially for clinical applications. Dictionary learning methods do not need large number of training data, nor requiring the registration between PET and CT/MR images, making it suitable for scenarios where CNN methods cannot be deployed. However, traditional patch-based dictionary learning methods still have challenges. During the sparse coding phase, the images are separated into numerous independent patches as the input, which inevitably ignores the global correlation within the image [19]. Furthermore, during the image reconstruction step, numerous independent patches are aggregated into an image, which results in the staircase artifacts near the patch boundaries [20]. Finally, computational cost is another common concern when exploiting dictionary learning methods.

Recently, with the demonstrated effectiveness of CNN methods [21]–[27] , more and more works start to combine the concepts from CNN with previously well-studied models/theories. Among them, convolutional sparse coding (CSC) [28]–[30] was recently proposed and provided a novel perspective on the original sparse coding theory. Compared with traditional dictionary learning and sparse coding methods, CSC decomposes the whole input signal as the convolution of $n$ filters and its corresponding feature maps [19]. There is no need to partition the image into independent patches. As a result, the aggregation procedure is avoided, reducing the artifacts while also accelerating the computational speed. It has

This work was supported in part by the U.S. National Institutes of Health under Grant R01AG052653. [†] indicates equal contributions. * indicates Corresponding Authors (li.quanzheng@mgh.harvard.edu, liuhf@zju.edu.cn).

N. Xie is with State Key Laboratory of Modern Optical Instrumentation, College of Optical Science and Engineering, Zhejiang University, Hangzhou, China and Department of Radiology, Massachusetts General Hospital and Harvard Medical School, Boston, USA.

K. Gong, N. Guo and Q. Li are with Department of Radiology, Massachusetts General Hospital and Harvard Medical School, Boston, USA.

Z. Qin and Z. Wu are with Department of Nuclear Medicine, First Hospital of Shanxi Medical University, Taiyuan, Shanxi.

H. Liu is with State Key Laboratory of Modern Optical Instrumentation, College of Optical Science and Engineering, Zhejiang University, Hangzhou, China.

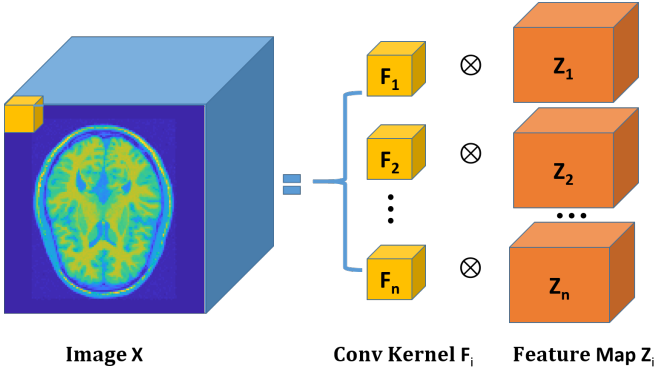

Fig. 1. The 3D convolution. Here the $3^{rd}$ dimension operates on the depth of the image $\mathbf{X}$, rather than the different channels (e.g. RGB channels) in traditional convolution.

been successfully applied to various image processing tasks, e.g. image super resolution [19], image fusion [31], 2D CT reconstruction [20] and MRI reconstruction [32][33].

In this work, we proposed a novel penalized-likelihood image reconstruction framework for PET using CSC. As clinical PET images are acquired and reconstructed in 3D mode, a 3D CSC framework was specifically developed, ensuring that the image structures across three dimensions can be jointly and efficiently recovered. Additionally, the 3D convolutional filters were pre-trained from the high-resolution 3D MR image, allowing the anatomical priors being introduced into PET image reconstruction. Finally, inspired by [34], we adopted a cube-based coding procedure without Fourier transform when solving the sparse pursuit problem. An 'ordered-subset' rule was developed to sparsely represent the residual image, instead of the original image, to further improve the computational efficiency.

The main contributions of this paper include: (1) a novel 3D CSC algorithm was designed to jointly incorporate structures along three dimensions. (2) To the best of our knowledge, this is the first work applying CSC to PET image reconstruction. (3) To further improve the computational efficiency, an 'ordered-subset' rule was adopted in the coding procedure and operated on the 3D residual image instead of the original image.

## II. METHOD

### A. PET Data Model

For PET imaging, given the measured data $\mathbf{y} \in \mathbb{R}^{M \times 1}$, which denotes the sum of the collected photons in PET detectors, the reconstruction procedure can be modeled through the affine transform as

$$\bar{\mathbf{y}} = \mathbf{G}\mathbf{x} + \mathbf{s} + \mathbf{r}, \qquad (1)$$

where $\bar{\mathbf{y}}$ denotes the expectation of $\mathbf{y}$, $\mathbf{x} \in \mathbb{R}^N$ represents the image ought to be recovered, $\mathbf{G} \in \mathbb{R}^{M \times N}$ is the system matrix, $\mathbf{s}$ and $\mathbf{r}$ are terms of scatter and random events, respectively, $M$ is the number of lines of response (LOR) and $N$ is the number of voxels in $\mathbf{x}$. Since $\mathbf{y}$ can be modeled by Poisson distribution, the likelihood function of $\mathbf{y}$ can be written as:

$$P(\mathbf{y}|\mathbf{x}) = \prod_{q=1}^{M} e^{-\bar{y}_q} \frac{\bar{y}_q^{y_q}}{y_q!}, \qquad (2)$$

where $q$ is the index of the detector pairs. The negative log-likelihood can correspondingly be written as

$$L(\mathbf{y}|\mathbf{x}) = \sum_{q=1}^{M} \bar{y}_q - y_q \log(\bar{y}_q) - \log(y_q!). \qquad (3)$$

### B. 3D-CSC Model

Zeiler et al. [28] proposed the CSC model in 2010. Unlike the traditional dictionary learning and the sparse coding methods, which partition the input image into independent patches, CSC operates and encodes the image by taking the whole input image into account. As a result, the consistency of different patches is exploited [19], and the staircase artifact inherited in the traditional patch-based sparse coding method can be alleviated. The typical 2D CSC model can be represented as:

$$\min_{z,f} \left\| \mathbf{x} - \sum_{i=1}^{n} \mathbf{f}_i * \mathbf{z}_i \right\|_F^2 + \lambda \sum_{i=1}^{n} \|\mathbf{z}_i\|_1, \qquad (4)$$

Here $\mathbf{x}$ denotes the 2D input image, $\{\mathbf{f}_i\}_{i=1,...,n}$ denotes the set of $s \times s$ sized convolutional filter, $\{\mathbf{z}_i\}_{i=1,...,n}$ is the set of feature maps with each $\mathbf{z}_i$ the same size as $\mathbf{x}$, and $*$ denotes the 2D convolution operator. In this model, image can be sparsely encoded by $\{\mathbf{f}_i * \mathbf{z}_i\}_{i=1,...,n}$.

With the wide application of 3D PET imaging, 2D CSC is not useful for PET reconstruction, as it cannot take advantage of information from the third dimension. In this work, we proposed a novel 3D CSC and implemented it for PET. The objective function is similar to the 2D CSC as

$$\min_{\mathbf{Z},\mathbf{F}} \left\| \mathbf{X} - \sum_{i=1}^{n} \mathbf{F}_i \otimes \mathbf{Z}_i \right\|_F^2 + \lambda \sum_{i=1}^{n} \|\mathbf{Z}_i\|_1. \qquad (5)$$

Here $\otimes$ denotes the 3D convolutional operator, $\{\mathbf{F}_i\}_{i=1,...,n}$ and $\{\mathbf{z}_i\}_{i=1,...,n}$ are 3D convolutional kernels and feature maps respectively, as shown in Fig.1. In this work, we pre-trained the 3D convolutional kernels $\{\mathbf{F}_i\}_{i=1,...,n}$ from the 3D MRI data. During the PET image reconstruction process, the learnt kernels were used to sparsely represent the 3D PET image. Through the kernels learnt from high-resolution MR images, the resolution of reconstructed PET images will be improved and the noise will also be reduced.

### C. Optimization

The 3D-CSC serves as the regularization term in our proposed reconstruction model as

$$\min_{\mathbf{x},\mathbf{Z}} L(\mathbf{y}|\mathbf{x}) + \beta \left( \left\| \triangle \mathbf{x} - \sum_{i=1}^{n} \mathbf{F}_i \otimes \mathbf{Z}_i \right\|_F^2 + \lambda \sum_{i=1}^{n} \|\mathbf{Z}_i\|_1 \right) \qquad (6)$$

Here $L(\mathbf{x}|\mathbf{y})$ is the negative log-likelihood defined in (3) which serves as the data fidelity term. The size of 3D convolutional filters $\mathbf{F}_i$ is $s \times s \times s$, and the 3D feature map $\mathbf{Z}_i$ has the same size with $\mathbf{x}$. $\beta$ and $\lambda$ are penalty parameters of the CSC and $l_1$ norm regularization, respectively. It is worth noting that instead of coding image $\mathbf{x}$, we code the residual image $\triangle \mathbf{x} = \mathbf{x} - \bar{\mathbf{x}}$, where $\bar{\mathbf{x}}$ denotes the mean image of $\mathbf{x}$ in the $s \times s \times s$ neighbored window, and the residual image $\triangle \mathbf{x}$

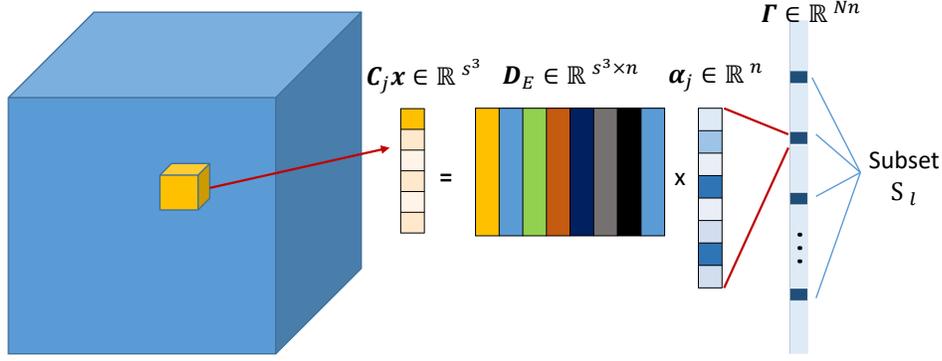

Fig. 2. The matrix multiplication form of 3D convolutional sparse coding, under the ordered-subset rule. In each batch of coding, we only update the $\boldsymbol{\alpha}_j$ where $j \in \mathrm{S}_l, \forall l$, which correspond to non-overlapped image cube $\boldsymbol{C}_j \boldsymbol{x}$ in 3D image $\boldsymbol{x}$.

---

**Algorithm 1:** 3D convolutional sparse coding for PET reconstruction

**Input:** Sinogram $\boldsymbol{y}$, system matrix $\boldsymbol{G}$, pre-trained elemental dictionary $\boldsymbol{D}_E$, penalty parameters $\beta$ and $\lambda$.

1: **Initialization:** For $k$=0, $\boldsymbol{x}^0 = \mathrm{FBP}(\boldsymbol{y})$, $\boldsymbol{\alpha}_j{}^0 =$
   $arg \min_{\boldsymbol{\alpha}_j} \left\| \frac{1}{n} \boldsymbol{C}_j \triangle \boldsymbol{x}^0 - \boldsymbol{D}_E \boldsymbol{\alpha}_j \right\|_F^2 + \lambda \|\boldsymbol{\alpha}_j\|_1$, $\forall j$

2: **For** $k$=1:*Maxit* **do**

3:    **For** iter=1: *Subit* **do**

4:       E-step: Compute $\Psi(\boldsymbol{x}; \boldsymbol{x}^k)$ by using (15)

5:       M-step: Solving the quadratic function (16) by using (17), update $\boldsymbol{x}^{k+1}$.

6:    **end**

7:    Compute $s$-cube neighbored mean image $\bar{\boldsymbol{x}}^{k+1}$, update $\triangle \boldsymbol{x}^{k+1} = \boldsymbol{x}^{k+1} - \bar{\boldsymbol{x}}^{k+1}$

8:    **For** $j \in \mathrm{S}_l, \forall l$ **do**

9:       Compute the residual $\boldsymbol{R}_i^{k+1} = \boldsymbol{C}_j \triangle \boldsymbol{x}^{k+1} - \sum_{\substack{p=1 \\ p \neq j}}^{N} \boldsymbol{D}_E \boldsymbol{\alpha}_j{}^k$.

10:       Sparse pursuit for (12): $\boldsymbol{\alpha}_j{}^{k+1} = arg \min_{\boldsymbol{\alpha}_j} \left\| \boldsymbol{R}_i^{k+1} - \boldsymbol{D}_E \boldsymbol{\alpha}_j \right\|_F^2 + \lambda \|\boldsymbol{\alpha}_j\|_1$

11:    **end**

12: **end**

13: **Output:** Reconstructed image $\boldsymbol{x} = \boldsymbol{x}^k$

---

represents the sparse texture of $\boldsymbol{x}$. Due to the replacement of $\triangle \boldsymbol{x}$, the convergence will be faster as demonstrated in our results shown later. To tackle the objective function (6), we divided it into 2 subproblems as

$$\boldsymbol{Z}_i{}^{k+1} = arg \min_{\boldsymbol{Z}_i} \left\| \left( \triangle \boldsymbol{x}^k - \sum_{\substack{p=1 \\ p \neq i}}^{n} \boldsymbol{F}_p \otimes \boldsymbol{Z}_p{}^k \right) - \boldsymbol{F}_i \otimes \boldsymbol{Z}_i \right\|_F^2 + \lambda \|\boldsymbol{Z}_i\|_1, \forall i. \quad (7)$$

$$\boldsymbol{x}^{k+1} = arg \min_{\boldsymbol{x}} L(\boldsymbol{y}|\boldsymbol{x}) + \beta \left\| \boldsymbol{x} - \bar{\boldsymbol{x}}^k - \sum_{i=1}^{n} \boldsymbol{F}_i \otimes \boldsymbol{Z}_i{}^{k+1} \right\| \quad (8)$$

Here we define $\boldsymbol{P}_i^k = \left( \triangle \boldsymbol{x}^k - \sum_{\substack{p=1 \\ p \neq i}}^{n} \boldsymbol{F}_j \otimes \boldsymbol{Z}_j{}^k \right)$ as the residue part of $\triangle \boldsymbol{x}^{k+1}$.

*1) Z Subproblem: Cube-based Coding.*

Traditional methods often update $\boldsymbol{Z}_i$ by converting the input signal to the Fourier domain [29][35][36]. Mathematically, this group of methods require treating the input as a whole signal, which undoubtedly will have computational and storage burdens, especially for 3D data. Inspired by [17][37], we proposed a 3D cube-based coding method to segment the 3D signal into groups of cubes and update them sequentially. Therefore, the signal will be divided into numerous subsets and can be coded in parallel. An essential contribution of [34][37] is that they provide a novel perspective which combines dictionary learning and the CSC. The signal can thus be broke down locally, making the coding procedure more straightforward.

In this work, the 3D convolution can be unfolded as the matrix multiplication, i.e. $\boldsymbol{x} = \boldsymbol{D}\boldsymbol{\Gamma}$. Here $\boldsymbol{D} \in \mathbb{R}^{N \times ns^3}$ is a band convolutional dictionary which is composed of the shifted elemental dictionary $\boldsymbol{D}_E \in \mathbb{R}^{s^3 \times n}$, while the rest of $\boldsymbol{D}$ is filled with zero entries. Each column of $\boldsymbol{D}_E$ is a vectorized 3D filter $\boldsymbol{F}_i, \forall i$. Correspondingly, $\boldsymbol{\Gamma} \in \mathbb{R}^{Nn}$ represents the vectorized $n$ feature maps. Accordingly, the traditional 3D CSC model (5) can be rewritten as:

$$\min_{\boldsymbol{D}, \boldsymbol{\Gamma}} \|\boldsymbol{x} - \boldsymbol{D}\boldsymbol{\Gamma}\|_F^2 + \lambda \|\boldsymbol{\Gamma}\|_1. \quad (9)$$

Now the optimization is more straightforward than that in Fourier-based methods. However, the computation and memory cost are still tremendous, as $\boldsymbol{\Gamma}$ is too large. For instance, if the size of 3D PET image is $100 \times 100 \times 100$ with 256 convolutional filters, the entry number of a single $\boldsymbol{\Gamma}$ is correspondingly 256 million. Under this circumstance, rather than directly updating the giant vector $\boldsymbol{\Gamma}$, we chose to update its fragment $\boldsymbol{\alpha}_j \in \mathbb{R}^n$, which corresponds to a $s \times s \times s$ sized cube in $\boldsymbol{x}$, i.e. $\boldsymbol{C}_j \boldsymbol{x} = \boldsymbol{D}_E \boldsymbol{\alpha}_j$, where $\boldsymbol{C}_j$ represents the operator of extracting cube at position $j$. The whole process is shown in Fig.2 and optimization of (9) can be reformulated as

$$\min_{\boldsymbol{D}_E, \boldsymbol{\alpha}} \left\| \boldsymbol{x} - \sum_{j=1}^{N} \boldsymbol{D}_E \boldsymbol{\alpha}_j \right\|_F^2 + \lambda \sum_{j=1}^{N} \| \boldsymbol{\alpha}_j \|_1. \tag{10}$$

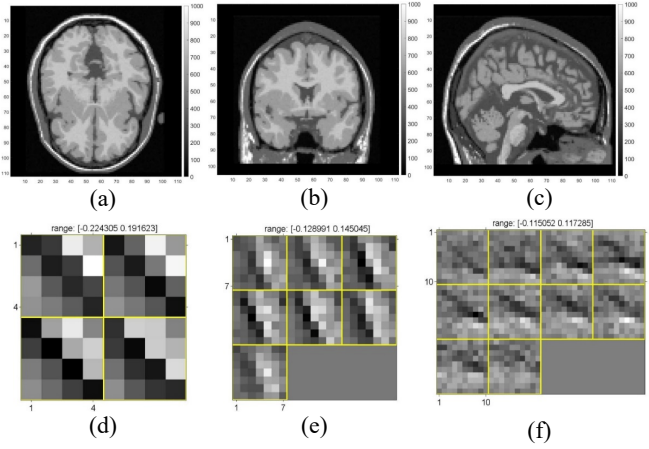

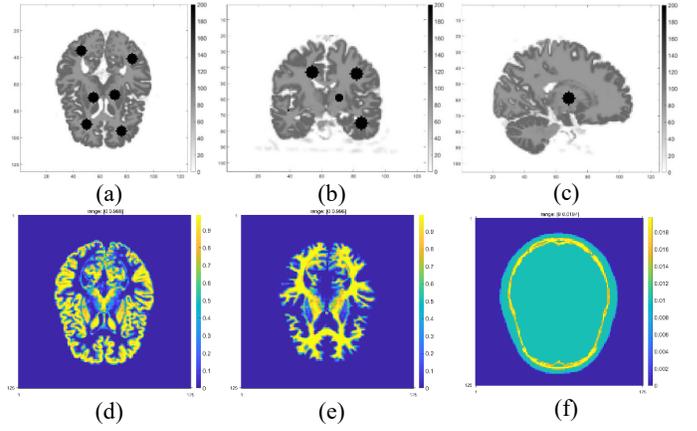

Fig. 3. The training for 3D convolutional filters. First row: brain MRI image slices from 3 orthogonal angles. Second row: Unfolded version for 24th filter cubes in 3 different cube size (d) $4 \times 4 \times 4$ (e) $7 \times 7 \times 7$ (f) $10 \times 10 \times 10$.

Fig. 4. The 3D simulated PET phantom. First row: 3 orthogonal slices of the 3D phantom. Here the black regions are the simulated tumor in this study. Second row: region maps for slice(a). (d) Gray matter (e) White matter (f) Attenuation map.

Here we pre-trained the elemental dictionary $\boldsymbol{D}_E$ from the 3D MRI data, as shown in Fig.3. Furthermore, for the purpose of further improving the computational efficiency, here we adopted an 'ordered-subset' rule to update $\boldsymbol{\alpha}_j$ in (10). We divided the voxel indexes into $s^3$ subsets,

$$S_l = \{l, l + s^3, l + 2 s^3, ..., N + l - s^3\}, l \\ = 1, 2, ..., s^3. \tag{11}$$

Similar to the ordered-subset (OS) mechanism for PET image reconstruction [38], the optimization of $\boldsymbol{\alpha}_j, j = 1, 2, ..., N$ has been successfully simplified to updating non-overlapping $\boldsymbol{\alpha}_j$ where $j \in S_l, \forall l$. Moreover, it also opens up a parallel computational way, as from 3D perspective, cubes in the same $S_l$ will not overlap with each other and thus not influence with each other. The detailed coding procedure is illustrated in Fig.2. Hence for our Z subproblem, (7) can be finally transformed to:

$$\boldsymbol{\alpha}_j^{k+1} = arg \min_{\boldsymbol{\alpha}_j} \left\| \boldsymbol{R}_i^k - \boldsymbol{D}_E \boldsymbol{\alpha}_j \right\|_F^2 + \lambda \| \boldsymbol{\alpha}_j \|_1, \\ j \in S_l, \forall l. \tag{12}$$

Here the residual $\boldsymbol{R}_i^k = \boldsymbol{C}_j \triangle \boldsymbol{x}^k - \sum_{p=1 \atop p \neq j}^{N} \boldsymbol{D}_E \boldsymbol{\alpha}_j^k$ serves as the input of this subproblem. In this sense, the update of the feature maps $\boldsymbol{Z}_i, \forall i$ has become sparse coding of $\boldsymbol{\alpha}_j$, where $j \in S_l, \forall l$. Also, for the initialization, we define the pursuit problem as follows,

$$\boldsymbol{\alpha}_j^0 = arg \min_{\boldsymbol{\alpha}_j} \left\| \frac{1}{n} \boldsymbol{C}_j \triangle \boldsymbol{x}^0 - \boldsymbol{D}_E \boldsymbol{\alpha}_j \right\|_F^2 + \lambda \| \boldsymbol{\alpha}_j \|_1, \\ j \in S_l, \forall l. \tag{13}$$

In this work, we used the least angle regression (LARS) algorithm [39] for the sparse pursuit problem in (12) and (13).

*2) X Subproblem*
Based on (1)(3) and (12), subproblem (8) can be rewritten as,

$$\boldsymbol{x}^{k+1} = arg \min_{\boldsymbol{x}} \sum_{q}^{M} \bar{y}_q - y_q \log(\bar{y}_q) + \\ \beta \left\| \boldsymbol{x} - \bar{\boldsymbol{x}}^k - \sum_{j=1}^{N} \boldsymbol{D}_E \boldsymbol{\alpha}_j^{k+1} \right\|_F^2 \tag{14}$$

where the constant term $\log(y_q!)$ is left out. Here we used the expectation maximization (EM) [40] to solve this problem.

***E-step***. We introduce $\hat{c}_{qj} = \frac{g_{qj} x_j^k}{\sum_j^N g_{qj} x_j^k + r_q + s_q} y_q$ as the expectation of the photons emitted from $j$-th voxel and also collected by $q$-th detector pair. Therefore, the voxel-wised optimization becomes

$$\boldsymbol{x}^{k+1} = arg \min_{\boldsymbol{x}} \Psi(\boldsymbol{x}; \boldsymbol{x}^k) \\ = arg \min_{\boldsymbol{x}} \sum_j^N \sum_q^N g_{qj} x_j - \hat{c}_{qj} \log(g_{qj} x_j) \\ + \beta \sum_j^N (x_j - \bar{x}_j^k - \left( \sum_{j=1}^{N} \boldsymbol{D}_E \boldsymbol{\alpha}_j^{k+1} \right)_j)^2. \tag{15}$$

***M-step***. Here we got a quadratic function by zeroing the following derivative:

$$\frac{\partial \Psi(\boldsymbol{x}; \boldsymbol{x}^k)}{\partial x_j} = 0 \iff A_j x_j + B_j + C_j \frac{1}{x_j} = 0 \tag{16}$$

And the solution is:

$$x_j^{k+1} = \frac{-B_j + \sqrt{B_j^2 - 4 A_j C_j}}{2 A_j}, \qquad A_j = 2\beta \\ B_j = \sum_q^M g_{qj} - \beta \bar{x}_j^k - \beta \left( \sum_{j=1}^{N} \boldsymbol{D}_E \boldsymbol{\alpha}_j^{k+1} \right)_j, \\ C_j = - \sum_q^M \frac{g_{qj} x_j^k}{\sum_j^N g_{qj} x_j^k + r_q + s_q} y_q. \tag{17}$$

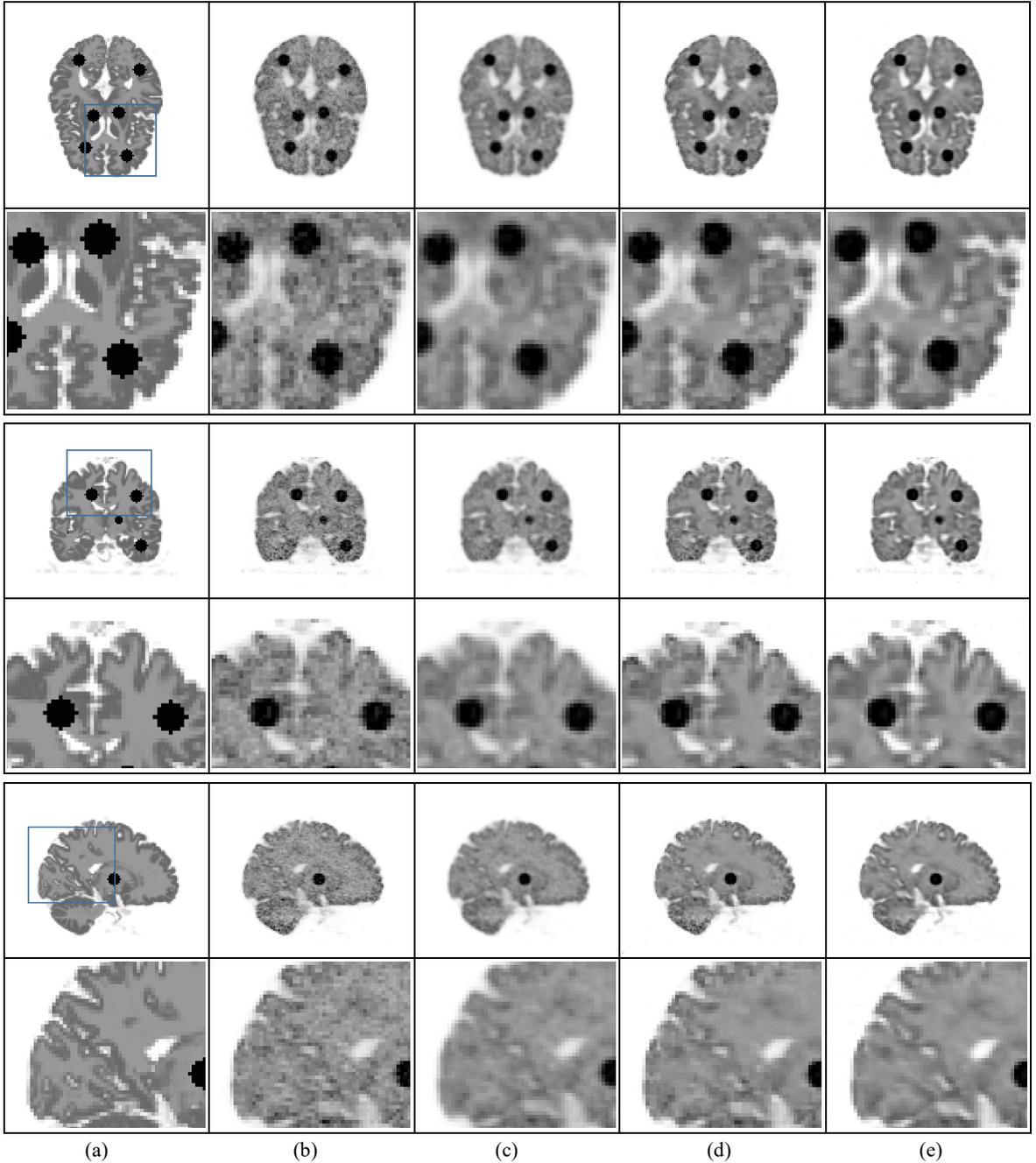

Fig. 5. Three orthogonal slices for high counted simulation data under $3 \times 10^8$ photon counts. (a) Ground truth (b) EM (c) Gaussian filtering (d) 3D Dictionary Learning (e) 3D PET-CSC

The overall procedure of the proposed method is presented in Algorithm 1. For the initialization, we adopted the FBP reconstruction as $\boldsymbol{x}^0$.

### D. Implementation

Our experiments were all conducted on Matlab R2013a in Linux system. The CPU model is Intel(R) Xeon(R) E5-4640, 2.40 GHZ.

#### 1) Filters Training

The 3D filter training is an essential part in our proposed method. We used a 3D $112 \times 112 \times 105$ sized brain MRI image to train the 3D convolutional filters, as shown in the first row of Fig.3. In addition, in order to demonstrate that our proposed method does not need registrations between PET and MR images, the MRI and PET data are different patients of the BrainWeb database [41].

The size of the convolutional filters also plays a key role in this work. As shown in Fig. 3, each filter can be translated into an elemental 3D structural feature. According to the figure, the larger cube size $s$ we chose, the more complex structure each 3D filter contained. Theoretically, larger convolutional cube provides larger receptive field, which contains more information. However, larger filter cubes also result in heavier computational cost, which is proportional to $s^3$ given the same filter number. As a tradeoff, we finally set the cube size $s$=4. As for the number of filters, we found that as long as the constructed filter group is over-completed, or the filter number

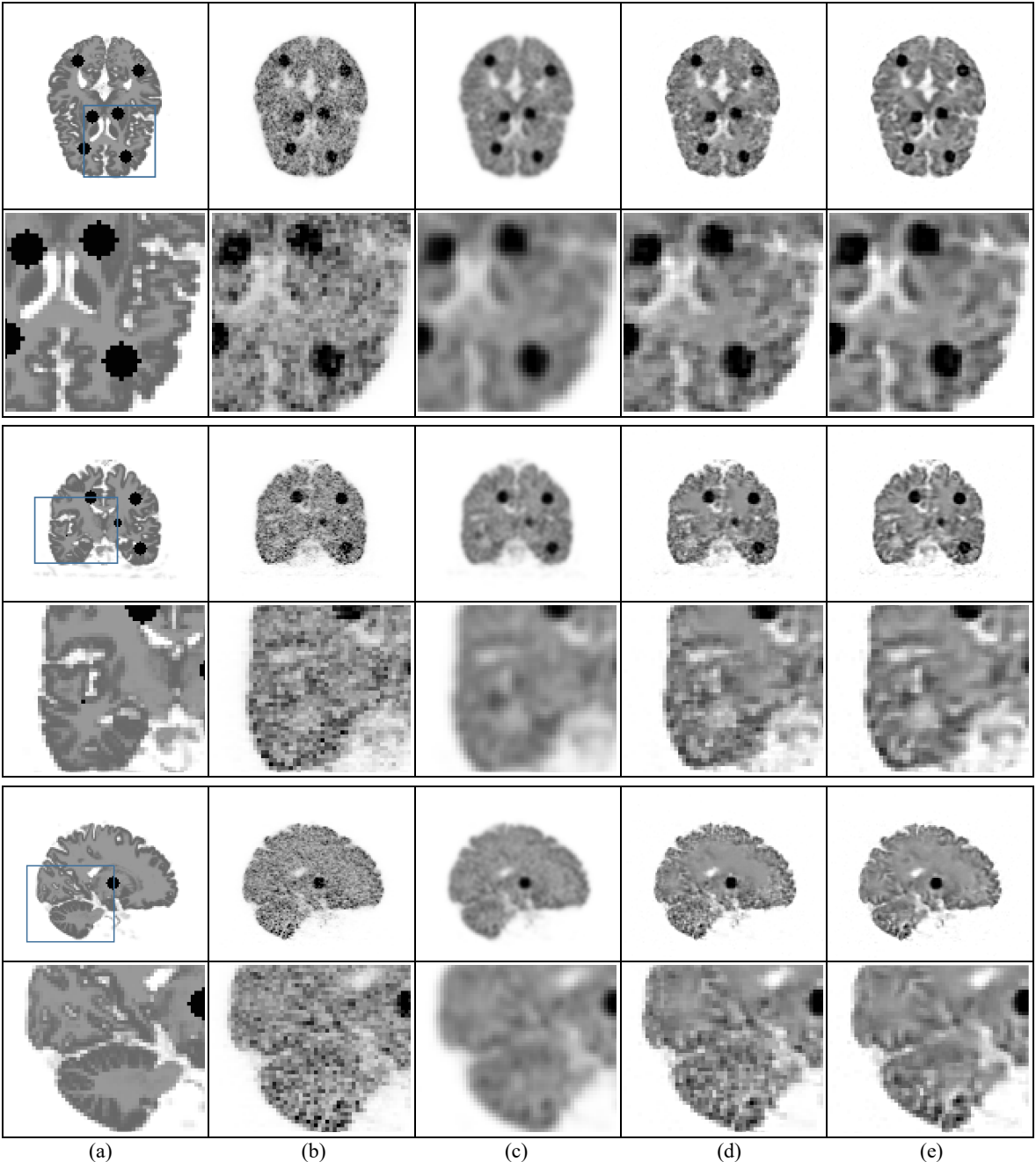

Fig. 6. Three orthogonal slices for high counted simulation data under $6 \times 10^7$ photon counts. (a) Ground truth (b) EM (c) Gaussian filtering (d) 3D Dictionary Learning (e) 3D PET-CSC

is larger than $s^3$ in other words, it is sufficient for 3D-CSC. In this case, we trained 256 filters, each with the size $4 \times 4 \times 4$.

*2) Penalty Parameters*

The penalty parameters $\beta$ and $\lambda$ are ought to be manually tuned. Here we have to admit that it is not practical to provide universal values for these two parameters, as their magnitudes should be varied according to different settings. In general, the magnitude of $\beta$ heavily depends on the choice of the system matrix $\boldsymbol{G}$ and the magnitude of $\lambda$ is influenced by the filter size and the range of image values. Nevertheless, there are still some empirical rules to follow. For $\lambda$, we introduced feature sparsity (*i.e.* the portion of non-zero entries in feature map) as the index for tuning. In our case, the sparsity is better to be tuned to be near $1.5 \times 10^{-3}$, while $\lambda$ is set to around 6.0. As for $\beta$, the magnitude ranges from $1 \times 10^{-9}$ to $1 \times 10^{-8}$ for our system matrix $\boldsymbol{G}$. In this study we adjusted $\beta$ for PET data under different counts. For instances we set $\beta = 1 \times 10^{-8}$ for $6 \times 10^7$ counted data and $\beta = 4 \times 10^{-9}$ for $3 \times 10^8$ counted data.

## III. EXPERIMENTS

In this part we conducted the experiments on both simulated data and clinical patient datasets. Also, three reference algorithms are included in this section for comparison.

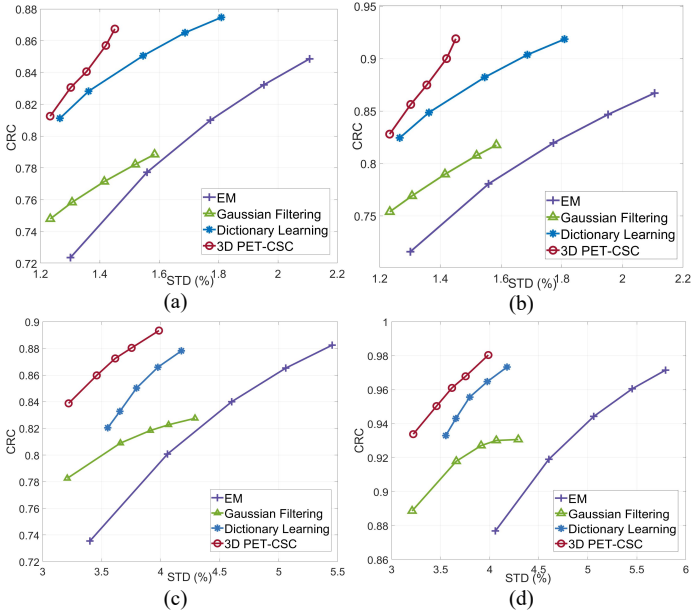

Fig. 7. CRC-STD curves for the reconstructions of high counted simulation data under $3 \times 10^8$ and $6 \times 10^7$ photon counts. (a) Tumor region for $3 \times 10^8$ counts data (b) Gray matter region for $3 \times 10^8$ counts data (c) Tumor region for $6 \times 10^7$ counts data (d) Gray matter region for $6 \times 10^7$ counts data

## A. Experimental Setup

### 1) Simulation data

In this simulation study, we adopted a 3D brain phantom from the BrainWeb database [41]. The image size is $125 \times 125 \times 105$ and the voxel size is $2.086 \times 2.086 \times 2.031 mm^3$. As we can see in Fig.4, the brain phantom includes the vessel region, gray matter region and white matter region. 12 tumors were inserted for quantification purposes. The geometry used in the simulation is based on the Siemens mCT scanner [42]. The tracer uptake is based on $^{18}$F-FDG. We generated the datasets of $6 \times 10^7$ and $3 \times 10^8$ photon counts to represent the low-count and high-count PET scenarios.

### 2) Clinical Data

In order to validate the proposed method for clinical scenarios, we employed $^{18}$F-FDG clinical brain and abdominal datasets in this study. All the data were acquired by the 5-ring GE Discovery MI PET/CT scanner. In the brain imaging study, the data of 30 min acquisition (30 min after FDG injection) was used. We down-sampled the 30-min dataset to 1/20 of its original count to generate ten low-count realizations. The ten realizations were reconstructed to quantify the noise performance. Moreover, we inserted a simulated tumor with diameter of 14 mm at the boundary of the white and gray matter for bias quantitative purposes. The reconstructed image size is $128 \times 128 \times 89$ and the voxel size is $2 \times 2 \times 2.8 mm^3$. For the abdominal dataset, we adopted a single frame scanned with duration of 180s. As demonstrated in Fig.9, we inserted 4 simulated tumors into the lung and liver regions. The tumors are of various uptakes and their diameters are 16.8mm, 15mm, 12mm and 9.6mm. The reconstructed image size is $128 \times 128 \times 89$ and the voxel size is $3 \times 3 \times 2.8 mm^3$.

### 3) Quantitative Evaluation

Contrast recovery coefficient (CRC) and standard deviation (STD) were used as the quantitative metrics. The CRC is defined as:

$$\text{CRC} = \frac{1}{R} \sum_{r=1}^{R} \frac{(\frac{\bar{a}_r}{\bar{b}_r} - 1)}{(\frac{a^{true}}{b^{true}} - 1)}. \tag{18}$$

Here $R = 10$ is the number of realizations we used, for both simulation data and real data. $\bar{a}_r = (1/K_a) \sum_{k=1}^{K_a} a_{r,k}$ is the average uptake for $K_a$ selected regions of interest (ROI) in $r$-th realization. For the simulation experiment, we respectively picked the ROIs in gray matter ($K_a = 10$) and tumor regions ($K_a = 12$). Similarly, $\bar{b}_r = (1/K_b) \sum_{k=1}^{K_b} b_{r,k}$ is the average value for $K_b$ background regions (white matter) in $r$th realization. Here we picked $K_b = 30$ background regions for simulation data. Correspondingly, the $a^{true}$ and $b^{true}$ are the ROI and background value in ground truth.

As for the STD, we computed it from the background as

$$\text{STD} = \frac{1}{K_b} \sum_{k=1}^{K_b} \frac{\sqrt{\frac{1}{R-1} \sum_{r=1}^{R} (b_{r,k} - \bar{b}_k)^2}}{\bar{b}_k}, \tag{19}$$

where $\bar{b}_k = (1/R) \sum_{r=1}^{R} b_{r,k}$ denotes the mean value of $k$ th background regions over $R$ realizations [43].

Given the fact that there is no ground truth for the real patient study, we alternatively adopted the contrast recovery ($\text{CR}_{tumor}$) for tumor regions,

$$\text{CR}_{tumor} = \frac{1}{R} \sum_{r=1}^{R} \bar{a}_r / a^{true}. \tag{20}$$

Similar with (18), $R$ is the number of realizations we reconstructed. Here we have $R = 10$ for the real brain study and $R = 1$ for the real body study. $\bar{a}_r = (1/K_a) \sum_{k=1}^{K_a} a_{r,k}$ is the average uptake for $K_a$ selected regions of interest (ROI) in $r$-th realization. Here we have $K_a = 1$ for the real brain study and $K_a = 4$ for the real body study. In addition, we employed the uptake ratio ($\text{UR}_{gray}$) for gray matter regions in real brain data:

$$\text{UR}_{gray} = \frac{1}{R} \sum_{r=1}^{R} \bar{a}_r / \bar{b}_r. \tag{21}$$

Here we respectively picked the ROIs in gray matter ($K_a = 10$) and $K_b = 10$ background regions for this calculation. Besides, for the single-realization abdominal dataset, we picked a background region in liver and computed its voxel-wised STD.

## B. Results

### 1) Simulation study

Fig.5 shows the slices of high-count ($3 \times 10^8$ photon counts) PET reconstruction from three orthogonal views. In this figure, we compared our proposed method with EM, EM plus Gaussian filtering, and 3D dictionary learning method. Here we set 10 iterations with 7 subsets for all algorithms. According to the figure, the proposed 3D CSC method shows superiority regarding edge preserving and denoising performance compared with other methods. More importantly, it successfully removes the staircase artifacts as shown in 3D dictionary learning method. Fig.7(a)(b) shows the CRC-STD curves for the $3 \times 10^8$ photon counts PET data. We can see the proposed method has lower STD and higher CRC on both tumor and gray matter ROIs. Fig.6 presents the slices of low-count ($6 \times 10^7$ photon counts) PET reconstruction from three orthogonal views. Fig.7(c)(d) present the CRC-STD curves for the PET data in $6 \times 10^7$ photon counts. For the low-count scenario, our proposed method still shows superior performance over other reference algorithms.

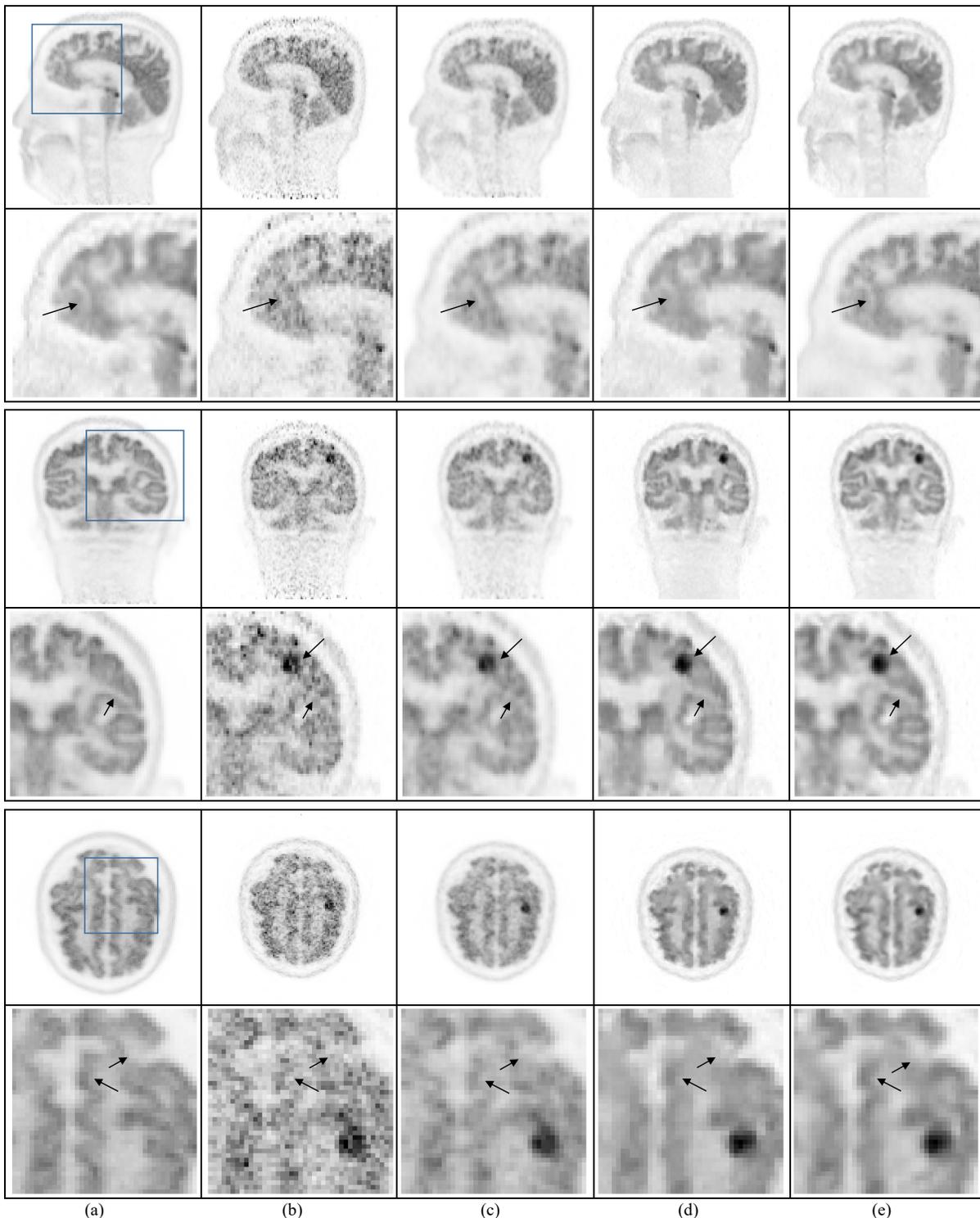

Fig. 8. Three orthogonal slices for real brain reconstruction. (a) High count data reconstructed by EM (b) EM (c) Gaussian filtering (d) 3D Dictionary Learning (e) 3D PET-CSC

### 2) Clinical Data

Fig.8 shows the slices of real brain data from three orthogonal views. According to the figure, our proposed method manages to recover more structures, such as tumor edge and cortex details, when compared with the dictionary learning method in the same noise level. As we can see from Fig.10, 3D

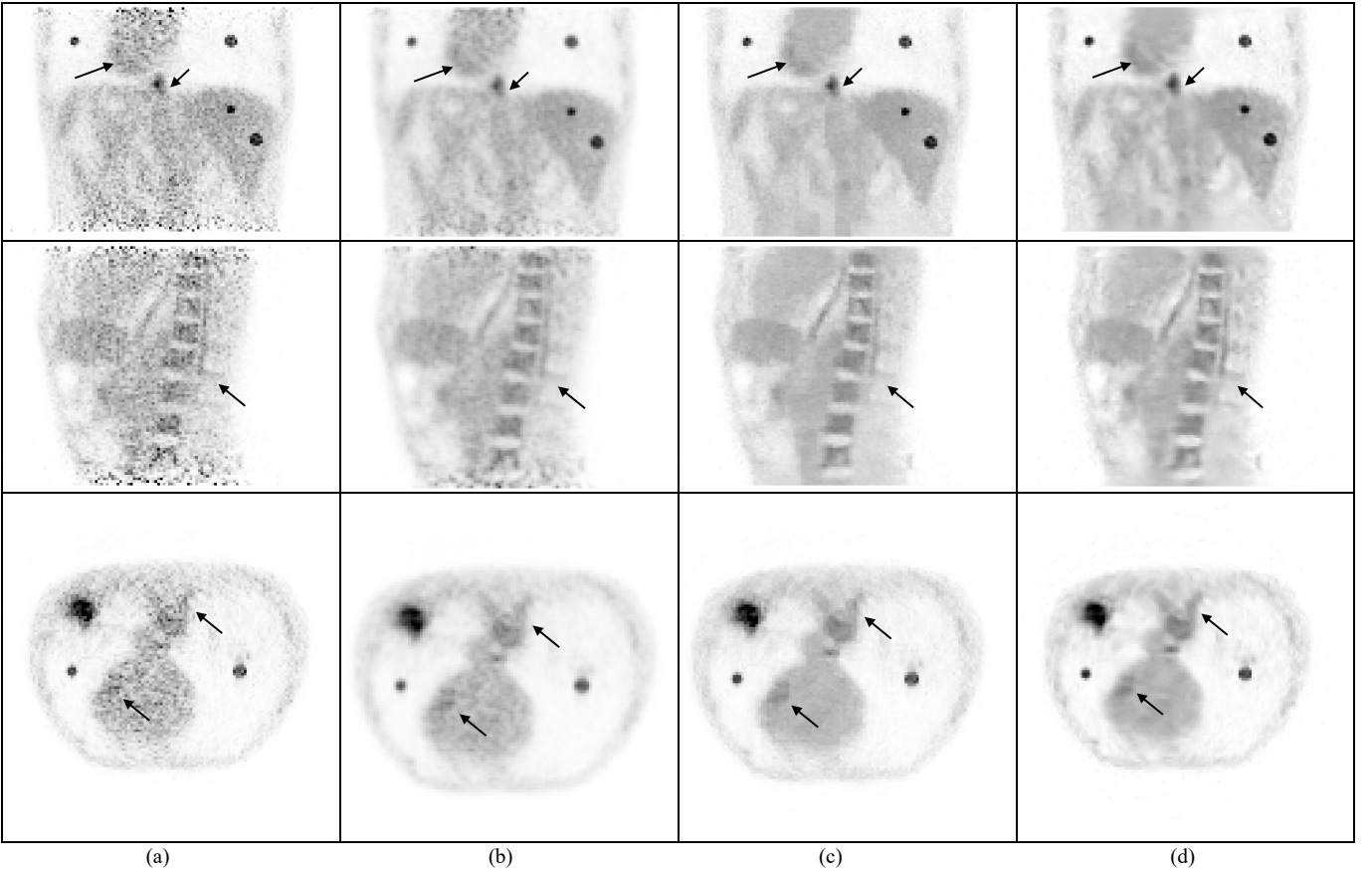

(a)              (b)              (c)              (d)

Fig. 9. Three orthogonal slices for real body reconstruction (a) EM (b) Gaussian filtering (c) 3D Dictionary Learning (d) 3D PET-CSC

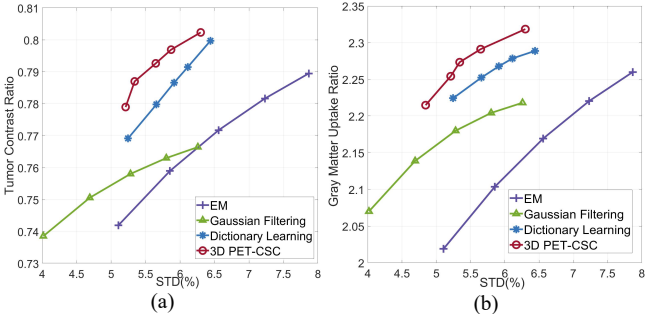

Fig. 10. $CR_{tumor}$-STD and $UR_{gray}$-STD curves for the reconstructions of real brain data. (a) Tumor region (b) Gray matter region

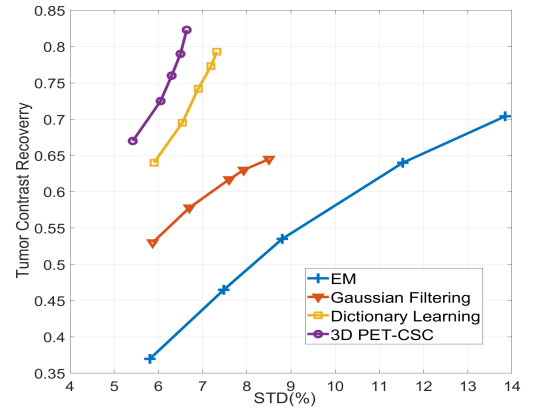

Fig. 11. The uni-realization $CR_{tumor}$-STD curves for real patient's body data.

PET-CSC is able to generate images with higher contrast on both tumor and gray matter regions, while reducing the image noise. Fig. 9 shows the reconstruction for the abdominal dataset. Our method can recover more structures and lower image noise when compared with other methods. This can be further demonstrated by the $CR_{tumor}$-STD curves shown in Fig.11.

### 3) Comparison with Traditional Sparse Coding

In Fig.12, we further explored our method regarding the convergence. As we mentioned in equation (6) of section 2.3, instead of coding the original 3D image $x$ during the image reconstruction, we coded the residual image $\triangle x = x - \overline{x}$ to further accelerate the convergence. In Fig.12(b), we compared the convergence speed of our 3D CSC with that of the traditional dictionary learning and sparse coding methods. In addition, we have also tested the convergent performance of our method when coding the original 3D image $x$. According to the figure, our method can substantially improve the convergent speed. Moreover, the introduction of residual image $\triangle x$ further improves the convergent performance. Besides, it can be seen that after the first iteration, the PSNR is already high. As demonstrated the sub-graph in Fig.12 (b), this can be credited to the introduction of the ordered-subsets coding mechanism in equation (12).

## IV. DISCUSSION

During the experiment set-up, we trained our 3D CSC filters from a 3D MRI image to utilize the high-resolution structure information. For our experiments, we chose the MRI images from another patient to demonstrate that our proposed method does not need the matching between PET and MR images. The main advantages of our work are threefold. The first advantage is that, unlike CNN methods, our 3D PET-CSC need no pre-training and no matching between PET and MR images. Secondly, 3D PET-CSC fully utilizes the information within the whole image, so that it can alleviate the staircase artifacts from the aggregation of patches. The third advantages is relevant to the convergence and computational efficiency. As shown in Fig.12, the proposed 3D CSC converges much faster than traditional sparse coding due to the mechanisms of "ordered-subsets" and "coding residual images". For our experiments, the average time cost of each iteration for our method is 23.6s, while for traditional sparse coding it is 574.3s.

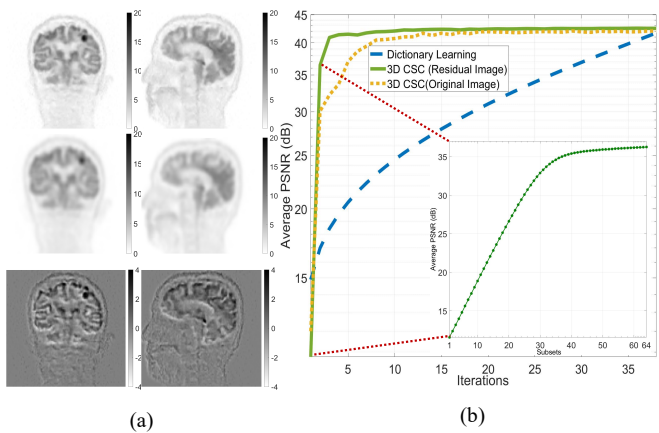

(a)                    (b)

Fig. 12. Convergence comparison for our proposed method and traditional dictionary learning method. (a) In each iteration, we handle the 3D residual image $\triangle x = x - \bar{x}$ instead. First row: the original image $x$. Second row: the mean image $\bar{x}$. Third row: the residual image $\triangle x$. (b) The average PSNR for different coding methods along with iterations. The sub-figure plots the PSNR along with the coded subsets in the $1^{st}$ iteration of proposed method.

In this work, we only employ the MR image during the filter learning phase and it is possible to include high-count PET data also in the training phase. This will be one of our future work. In addition, for our proposed framework, there are two parameters, $\beta$ and $\lambda$, to adjust. In this work, we only choose an empirical rule to adjust those parameters. More elegant parameter adjusting protocols are needed and worth exploring.

## V. Conclusion

We have proposed a novel 3D CSC regularization into 3D PET image reconstruction. The 3D structure CSC method was developed and incorporated into the PET reconstruction model for the first time. We have also applied a framework for coding 3D residual image with an ordered-subset mechanism. Based on the simulation and clinical studies, we can see that the proposed 3D PET-CSC presents an efficient and robust approach to incorporate the anatomical prior into PET reconstruction without the need for registration and large number of datasets. Compared with traditional dictionary learning and sparse coding methods, the proposed method manages to alleviate the staircase artifact and recover the image with superior quality.